\documentclass[%
preprintnumbers,
nofootinbib,
 amsmath,amssymb,
 aps, physrev,
 twocolumn]{revtex4-2}
\usepackage{amsmath}
\usepackage{graphicx}
\usepackage{float}
\usepackage{dcolumn}
\usepackage{bm}
\usepackage{hyperref}
\usepackage{multirow}
\usepackage{xspace}
\usepackage{xcolor}
\usepackage{aas_macros}

\newcommand{\SMW}{\texttt{SMW}\xspace}
\newcommand{\MWest}{\texttt{MWest}\xspace}

\newcommand{\MilkyWayest}{\texttt{MilkyWay-est}\xspace}
\newcommand{\Symphony}{\texttt{Symphony}\xspace}
\newcommand{\BabyBison}{\textsc{BabyBison}\xspace}
\newcommand{\bbbar}{$b\bar{b}$\xspace}
\newcommand{\tptm}{$\tau^+\tau^-$\xspace}

\begin{document}
\preprint{FERMILAB-PUB-26-0839-PPD-T}

\title{The Sensitivity of Gamma-Ray Searches for Dark Matter Annihilation in \\ Milky Way Dwarf Galaxies: the LSST Era and Beyond}

\author{Ilumi Chavez}
\affiliation{
Department of Physics and Astronomy, University of North Florida, Jacksonville, Florida 32224, USA
}
\author{Aditya Ghosh}
\affiliation{
Department of Physics and Astronomy, University of North Florida, Jacksonville, Florida 32224, USA
}
\author{Chris Kelso}
\affiliation{
Department of Physics and Astronomy, University of North Florida, Jacksonville, Florida 32224, USA
}
\author{Dan Hooper}
\affiliation{Department of Physics, University of Wisconsin, Madison, WI 53706, USA}
\affiliation{Wisconsin IceCube Particle Astrophysics Center (WIPAC),
University of Wisconsin, Madison, WI 53703, USA}
\author{Alex Drlica-Wagner}
\affiliation{Fermi National Accelerator Laboratory, P.O. Box 500, Batavia, IL 60510, USA}
\affiliation{Kavli Institute for Cosmological Physics, University of Chicago, Chicago, IL 60637, USA}
\affiliation{Department of Astronomy and Astrophysics, University of Chicago, Chicago, IL 60637, USA}

\author{Ethan~O.~Nadler}
\affiliation{Department of Astronomy \& Astrophysics, University of California, San Diego, La Jolla, CA 92093, USA}

\date{\today}
\begin{abstract}

We forecast the sensitivity of gamma-ray searches to dark matter annihilation in Milky Way satellite dwarf galaxies using \BabyBison, a Monte Carlo pipeline combining cosmological satellite simulations and observational selection functions with  a joint likelihood analysis of gamma-ray data. Although LSST is expected to substantially expand the catalog of known satellite galaxies, we find that these discoveries are likely to produce only modest improvements in the resulting dark matter annihilation constraints, typically below 20\%. Much more substantial gains are possible, however, in relatively rare realizations that contain yet-to-be-discovered satellites with large $J$-factors. These results reinforce the fact that indirect detection analyses using the population of dwarf galaxies are typically driven in large part by a small number of high-$J$-factor systems rather than by the total satellite count. We present projections for the sensitivity of Fermi-LAT and the proposed Advanced Particle-astrophysics Telescope (APT) to dark matter annihilation, finding that a complete census of the Milky Way dwarf sample could enable such instruments to probe dark matter with masses up to the TeV scale. We also investigate the dark matter discovery potential of these instruments and project their ability to measure the properties of an annihilating dark matter particle.

\end{abstract}

\maketitle

\section{Introduction}
\label{sec:intro}

The nature of dark matter remains one of the most intriguing unsolved problems in contemporary physics. Experiments on Earth and in space have long sought to detect dark matter particles through their non-gravitational interactions. Among the most widely studied dark matter candidates are thermal relics of the early universe~\cite{Bertone:2016nfn}.  Particles with weak-scale masses and couplings can freeze out of thermal equilibrium with a relic abundance consistent with the measured dark matter density --- a fact often referred to as the ``WIMP miracle.'' This observation motivates our work and has driven extensive interest in weakly interacting massive particles (WIMPs), which arise in a variety of theories beyond the Standard Model of particle physics.

For a self-conjugate thermal relic with velocity-independent annihilations, the observed dark matter abundance is obtained for a thermally averaged annihilation cross section of approximately $\langle\sigma v\rangle \simeq 2.2\times 10^{-26}\,\mathrm{cm^{3}\,s^{-1}}$ over much of the weak-scale mass range~\cite{Steigman:2012nb}. This benchmark motivates indirect searches for dark matter's annihilation products, including photons, neutrinos, electrons, positrons, antiprotons, and heavier antinuclei. Such searches target dark matter in the Galactic Center, dwarf galaxies, galaxy clusters, and the broader Galactic halo, as well as dark matter particles gravitationally captured in the Sun (for a review, see Ref.~\cite{Slatyer:2021qgc}).

Milky Way satellite dwarf spheroidal galaxies are particularly promising targets for gamma-ray searches because they are dark matter dominated and produce little intrinsic astrophysical gamma-ray emission. Combining observations of multiple dwarfs in a joint likelihood analysis can further improve sensitivity. Searches with the Fermi Large Area Telescope (Fermi-LAT) have placed stringent upper limits on the dark matter annihilation cross section in the absence of a statistically significant signal. For annihilations to $b\bar{b}$, for example, published limits exclude the thermal-relic benchmark below masses of approximately $m_\chi \sim 50$--$100\,\mathrm{GeV}$, depending on the dwarf sample, adopted $J$-factors, and other analysis assumptions~\citep{Albert2017,DiMauro2022,McDaniel:2023bju,Circiello:2026dSph,Bertolez-Martinez:2026yqt}. It should be noted that these studies are subject to a number of observational and modeling systematics, including uncertainties associated with stellar membership~\cite{2016MNRAS.462..223B,Ichikawa:2017rph}, unresolved binaries~\cite{Minor_2019}, departures from equilibrium~\cite{Tchiorniy_2026}, departures from spherical symmetry~\cite{Bonnivard:2014kza,Hayashi:2016kcy,Klop:2016lug}, and assumptions regarding the stellar~\cite{Bonnivard:2014kza,Chiappo:2018mlt} and dark matter~\cite{Klop:2016lug} density profiles. Such factors could cause the resulting constraints to be overly stringent by a factor of up to a few~\cite{Bertolez-Martinez:2026yqt}. Modest gamma-ray excesses have been reported toward individual systems, including Reticulum II, Sculptor, and Tucana II, but neither individual dwarfs nor combined analyses have thus far yielded compelling evidence for dark matter annihilation~\citep{DiMauro2022,McDaniel:2023bju}.

Existing joint-likelihood analyses of Fermi-LAT data have considered samples of approximately 50 known Milky Way dwarf galaxies. The Vera C.~Rubin Observatory's Legacy Survey of Space and Time (LSST) is expected to discover dozens of new Milky Way satellites~\citep{Drlica-Wagner2019,LSSTDESC:2025hol}.  Recent forecasts combine realistic models of the survey sensitivity and source classification with satellite populations derived from dark-matter-only cosmological zoom-in simulations and empirical models of the galaxy--halo connection. These discoveries could improve the sensitivity of stacked gamma-ray searches, particularly if they include nearby systems with large $J$-factors. Quantifying this anticipated improvement requires modeling both the underlying satellite population and the observational selection that determines which systems enter the analysis.

In this study, we introduce \textsc{BabyBison}\footnote{\href{https://github.com/carlinist/Bison_Farm}{\textbf{https://github.com/carlinist/Bison\_Farm}}}, a parallelized Monte Carlo pipeline that forecasts sensitivity to dark matter annihilation through a likelihood analysis of dwarf galaxies. The pipeline accommodates both observed dwarf galaxy catalogs and large simulated satellite populations. We use it to evaluate (i) the projected sensitivity of satellite samples available before and after LSST, (ii) the \emph{ultimate} sensitivity attainable with a complete census of the modeled Milky Way dwarf galaxy population, and (iii) the precision with which the dark matter mass and annihilation cross section could be recovered if a signal was detected.

The remainder of this paper is organized as follows.  In Sec.~\ref{sec:simulations}, we describe the dark-matter-only cosmological simulations, the model connecting subhalos to satellite galaxies, and the observational selection functions employed in our work.  In Sec.~\ref{sec:MC}, we describe our Monte Carlo analysis, and in Sec.~\ref{sec:results} we present our results.  We summarize our findings in Sec.~\ref{sec:conc}.

\section{Simulations and Satellite Model}
\label{sec:simulations}

The analysis presented here is based on dark-matter-only cosmological zoom-in simulations of Milky Way-mass host halos from the \texttt{SymphonyMilkyWay} (\texttt{SMW})~\citep{Mao2015SubhaloConcentration, Nadler2023Symphony} and \texttt{MilkyWay-est} (\texttt{MWest})~\citep{Buch2024} suites. The \SMW suite is part of the \Symphony compilation of 262 simulations spanning host halo
masses of $10^{11}$--$10^{15}\,M_\odot$~\citep{Nadler2023Symphony}. The full \SMW suite contains 45
Milky Way-mass host halos originally simulated by~\citet{Mao2015SubhaloConcentration} and selected from the
\texttt{c125-1024} parent simulation. The hosts each have a characteristic
present-day virial mass of
$M_{\mathrm{host}}=10^{12.09\pm0.02}\,M_\odot$, defined according to the \citet{Bryan:1997dn} definition with an overdensity parameter $\Delta_{\mathrm{vir}}\approx 99$ times the critical density of the universe today.\footnote{We adopt cosmological parameters of $h = 0.7$, $\Omega_m = 0.286$,
$\Omega_{\Lambda} = 0.714$, $\sigma_8 = 0.82$, and $n_s = 0.96$, consistent with the \SMW and \MWest suites.}
The high-resolution dark-matter particles have a mass of
$m_{\mathrm{part}}=4.0\times10^{5}\,M_\odot$, and the simulations
have a comoving Plummer-equivalent force-softening scale of
$\epsilon=170\,\mathrm{pc}\,h^{-1}$.
The minimum well-resolved subhalo virial mass is $1.2\times10^{8}\,M_\odot$, corresponding to
approximately $300\,m_{\mathrm{part}}$ and a characteristic
subhalo-to-host mass ratio of
$M_{\mathrm{sub}}/M_{\mathrm{host}}\simeq1.1\times10^{-4}$~\citep{Nadler2023Symphony}. 

Following Ref.~\cite{DES:2019ltu}, we use the two halos
from the full 45-host \SMW suite that feature Large Magellanic Cloud
(LMC) analogs with realistic internal and orbital properties. These systems also experienced an early major merger resembling the Gaia-Sausage-Enceladus event. The galaxy--halo connection model described below was fit to the Milky Way satellite population using these two simulations. Our predictions based on these two \SMW hosts are therefore calibrated to the observed satellite population and extend to the yet-to-be-discovered population of dwarf galaxies. However, because this subset contains only two host halos with specific assembly histories and LMC analog properties, it cannot fully characterize the host-to-host scatter in the predicted Milky Way satellite populations.

\begin{figure*}
    \centering
    \includegraphics[width=0.95\textwidth]{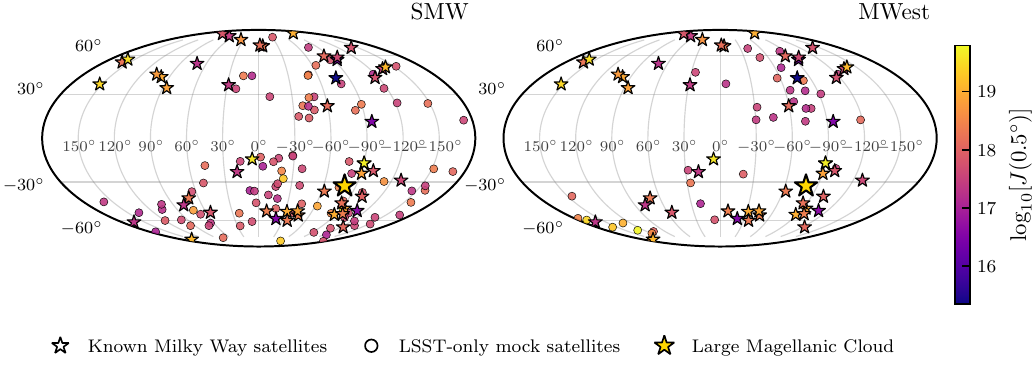}
    \caption{The distribution in Galactic coordinates of currently known Milky Way satellites (stars) and dwarf galaxies predicted to be discovered by LSST (circles), shown for a single representative Monte Carlo realization from each of the \SMW and \MWest simulation suites. The dwarf galaxies are colored according to their assigned $J$-factors for the mock satellites and measured $J$ factors for the known satellites (in units of ${\rm GeV}^2/{\rm cm}^5$). Note that dwarf galaxies at $|b| < 10^{\circ}$ are excluded by the Galactic-plane cut applied throughout our analysis.  The LMC is indicated by a large star.} 
    \label{fig:gal_plot}
\end{figure*}

To better estimate the expected host-to-host scatter, we use \MWest, a suite of 20 simulations selected to reproduce key aspects of the Milky Way's assembly
history~\citep{Buch2024}. Candidate hosts were selected from the
\texttt{c125-1024} parent simulation to have Milky Way-like masses and
concentrations, a recently accreted LMC analog on a realistic orbit, and an
earlier Gaia-Sausage-Enceladus-like major merger. We use all 20 \MWest main halos. We analyze each \MWest host at the snapshot when its LMC analog is closest to a distance of 50 kpc from the center of the MW halo. At these snapshots, the MW host halos have virial masses of $M_{\mathrm{host}}\simeq10^{12.10\pm0.08}\,M_\odot$.
The \MWest simulations use the same numerical resolution as \SMW. We adopt a conservative common minimum subhalo-to-host mass ratio of
$M_{\mathrm{sub}}/M_{\mathrm{host}}>1.4\times10^{-4}$, which corresponds to
$300\,m_{\mathrm{part}}$ for the lowest-mass \MWest host~\citep{Buch2024}. 

We impose the same resolution cuts on simulated subhalos from both \SMW and \MWest. We require a peak halo mass of $M_{\rm peak}>10^7\,M_\odot$, a peak maximum circular velocity of $V_{\rm peak}>10\,\mathrm{km\,s^{-1}}$, and a maximum circular velocity of $V_{\rm max}>9\,\mathrm{km\,s^{-1}}$. Orphan satellites are treated separately and are required to have $M_{\rm peak}>10^7\,M_\odot$ and a Galactocentric radius smaller than $300\,\mathrm{kpc}$. Additional selections were placed on simulated satellite galaxy size, luminosity, and distance when applying the observational selection functions described in Sec.~\ref{sec:selfunc}.

We did not refit the galaxy--halo connection model to the \MWest simulations. Rather, we directly applied the model that was fit using two selected \SMW hosts (see Section \ref{sec:ghm}).
Thus, our \MWest predictions should be treated as rough estimates of the total Milky Way satellite population, although we expect them to reasonably characterize the expected degree of host-to-host variance.

Our Monte Carlo runs use stratified sampling, generating an equal number of realizations for each host halo.  We use these simulated subhalo catalogs as inputs to a subhalo-satellite connection model that includes prescriptions for baryonic disruption and orphan treatment to generate populations of dwarf galaxies~\cite{Nadler2019,Nadler2020}.  For each realization, we randomly select an observer position from the vertices of an octahedron centered on the host halo, with each vertex located $8.5 \, {\rm kpc}$ from the halo center. We then apply survey selection functions to the simulated satellite population~\cite{LSSTDESC:2025hol} and assign $J$-factors~\cite{PaceStrigari2019} for use in the subsequent indirect detection analysis.  Each of these steps is described in more detail in the following sections.

\begin{figure*}[t]
    \centering
    \includegraphics[width=0.95\textwidth]{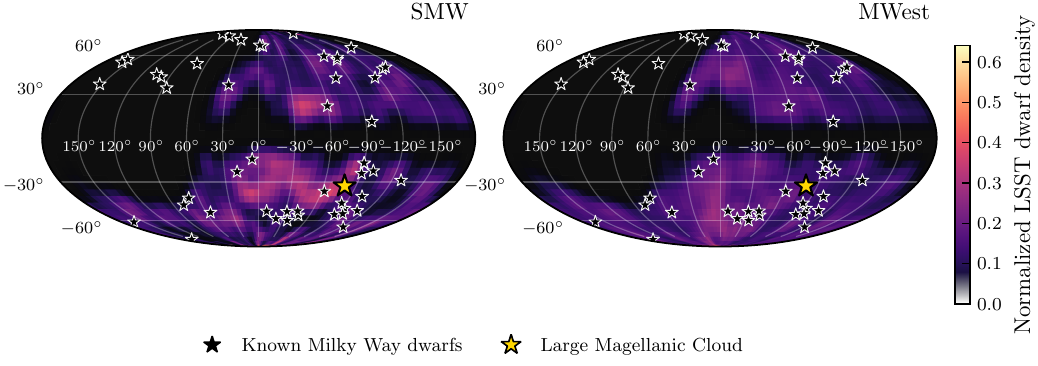}
    \caption{The distribution in Galactic coordinates of the dwarf galaxies projected to be discovered by LSST in the \SMW and \MWest simulation suites. The heat map shows the normalized density per solid angle of dwarf galaxies predicted to be discovered by LSST, stacked over the 30{,}000 (\SMW) and (\MWest) Monte Carlo realizations. Again, the $|b| < 10^{\circ}$ region has been masked in our analysis. Black stars (outlined in white) mark the locations of currently known dwarf galaxies. Both suites predict anisotropic discovery patterns for LSST, with enhanced densities in the vicinity of the LMC.}
    \label{fig:lsst_probability_density_current_known}
\end{figure*}

\subsection{Subhalo-Satellite Connection}
\label{sec:ghm}

We utilize the empirical framework developed in Refs.~\cite{Nadler2019,Nadler2020} to populate the subhalos in each simulation suite with satellite galaxies. We do not refit the galaxy--halo connection model in this work. Instead, we sample its parameters from the joint posterior distribution inferred in Ref.~\cite{Nadler2020} by fitting the Milky Way satellite populations observed in the Dark Energy Survey (DES) \citep{DES:2016jjg} and the Pan-STARRS1 survey (PS1) \citep{PS1:2016}. Our realizations therefore incorporate the uncertainty in the galaxy--halo connection model in addition to the host-to-host variation present in the simulated halo populations. For systems with $M_V<-13\,\mathrm{mag}$, satellite luminosities are assigned through an abundance-matching relation calibrated to the GAMA galaxy survey~\cite{10.1093/mnras/stv1013} and based on the largest maximum circular velocity attained by a subhalo. This relation is extrapolated to fainter satellites using a faint-end slope and intrinsic scatter sampled from the posterior distribution inferred in Ref.~\cite{Nadler2020}. This model assumes Gaussian scatter in absolute magnitude that is independent of the properties of the halo.  

Satellite half-light radii, $r_{1/2}$, are drawn from a lognormal distribution with a mean that is a function of the subhalo's virial radius at accretion, and a scatter that is a free parameter fit by the model.  We require $r_{1/2}>10~\mathrm{pc}$ to exclude likely star clusters from the modeled population.

To account for enhanced tidal stripping of subhalos due to the Galactic disk, we apply the baryonic disruption model of Ref.~\cite{Nadler2018} to the dark matter--only simulations. This model uses a random forest classifier trained on hydrodynamic simulations to assign each subhalo a disruption probability, $p_{\mathrm{disrupt}}$, based on its orbital and accretion properties. The classifier uses five inputs: the distance from the host halo center and the scale factor at pericenter, and the subhalo mass, maximum circular velocity, and scale factor at accretion. This correction is particularly important in the inner halo, where baryonic effects most efficiently destroy subhalos.

We employ the galaxy occupation fraction parametrization of Ref.~\cite{Nadler2019}, which smoothly suppresses galaxy formation in low-mass halos:
\begin{equation}
    f_{\mathrm{gal}}(M_{\mathrm{peak}}) =
    \frac{1}{2}\left[1+\mathrm{erf}\!\left(
    \frac{\log M_{\mathrm{peak}}-\log M_{50}}
    {\sqrt{2}\,\sigma_{\mathrm{gal}}}
    \right)\right],
\end{equation}
where $M_{\mathrm{peak}}$ is the largest virial mass attained by a subhalo, $M_{50}$ is the peak halo mass at which 50\% of halos host galaxies, and $\sigma_{\mathrm{gal}}$ sets the width of the transition in logarithmic halo mass. $M_{50}$ and $\sigma_{\mathrm{gal}}$ are each free parameters fit by the model.  

Following~\citet{Nadler2019}, we include orphan satellites to account for galaxies whose host subhalos are lost because of numerical resolution rather than true physical disruption. We adopt model parameters that treat subhalo disruption in hydrodynamical simulations as physical and disruption in dark-matter-only simulations as a numerical effect.  Orphan systems are particularly important in modeling the faint satellite population near the resolution limit.

In Fig.~\ref{fig:gal_plot}, we show a single representative realization from each simulation suite. For each realization, we rotate the satellite system so that the simulated LMC analog lies at the observed sky position of the LMC, while preserving its simulated distance. Black stars mark the currently known Milky Way satellites, and colored circles denote simulated systems that would be discovered by LSST but are not included in the classical, DES, or PS1 satellite samples~\cite{LSSTDESC:2025hol}.  The stars and circles are colored by their assigned $J$-factors. The empty band around the Galactic plane reflects the $|b| > 10^{\circ}$ cut applied throughout our analysis to avoid this astrophysically complex region of the sky. The distribution also reflects LSST's predominantly southern equatorial footprint.

In Fig.~\ref{fig:lsst_probability_density_current_known}, we show the normalized sky density of these LSST-only systems, combined across all 30{,}000 \texttt{SMW} and \texttt{MWest} realizations. Both suites predict anisotropic distributions of new discoveries, with higher densities far from the Galactic plane and near the LMC, as previously noted in Ref.~\cite{DES:2019ltu}. This anisotropy also reflects the predominantly southern footprint of the LSST survey, which restricts the regions in which new satellites can be detected. The \texttt{SMW} distribution exhibits more pronounced localized enhancements, whereas the \texttt{MWest} distribution is smoother across the same regions. 

On a per-realization basis, \SMW predicts that LSST will discover a median of $66^{+20}_{-16}$ dwarf galaxies compared to $45^{+16}_{-11}$ for \MWest, where the uncertainties indicate the central 68\% intervals. The characteristic properties of the predicted satellites are similar between the two suites. The median projected half-light radii are $r_{1/2}=61^{+26}_{-16}\,\mathrm{pc}$ for \SMW and $r_{1/2}=65^{+31}_{-18}\,\mathrm{pc}$ for \MWest. The median absolute magnitudes are also similar, with $M_V=-3.13^{+0.34}_{-0.40}$ for \SMW and $M_V=-3.06^{+0.42}_{-0.47}$ for \MWest, while the median heliocentric distances are $138^{+10}_{-12}\,\mathrm{kpc}$ and $140^{+18}_{-17}\,\mathrm{kpc}$, respectively. These similarities suggest that the smoother distribution of satellites seen in \MWest (Fig.~\ref{fig:lsst_probability_density_current_known})  is not driven by systematic differences in the characteristic sizes, luminosities, or distances of the predicted satellites. Instead, we attribute it primarily to averaging over the diverse satellite distributions of 20 independent \MWest hosts, which smooths out localized structures that remain visible in the two-host \SMW sample.

\subsection{Observational Selection Function}
\label{sec:selfunc}

We apply survey-dependent observational selection functions to create mock observations of the Milky Way satellite galaxy population. 
Following Ref.~\cite{Nadler2020}, we compute each satellite's detection probability, $p_{\mathrm{detect}}$, from its luminosity, projected half-light radius, heliocentric distance, and sky position, accounting for the survey footprint and depth. The overall probability that a satellite is included in a given survey catalog is $p_{\mathrm{survey}}=p_{\mathrm{detect}}\left(1-p_{\mathrm{disrupt}}\right)f_{\mathrm{gal}}$.

We construct three catalogs for each realization, corresponding to different observational scenarios. The pre-LSST catalog includes classical dwarf galaxies, defined as systems with $M_V \leq -8.8$, together with satellites detected by either DES or PS1.  The post-LSST catalog additionally includes satellites predicted to be detected by LSST, assuming idealized star-galaxy separation~\cite{LSSTDESC:2025hol}. The ``ultimate'' catalog assumes that the full population of Milky Way satellites has been observed. Because these catalogs are drawn from the same realization, they allow us to isolate the effects of survey selection while holding the underlying satellite population fixed. When generating mock observational samples for the pre- and post-LSST scenarios, we apply observational selection cuts requiring $M_V < 0$, $r_{1/2}>10~\mathrm{pc}$, and $D<300~\mathrm{kpc}$ as described in \citet{Nadler2020}. These cuts are intended to restrict the mock satellite samples to the regime where dark-matter-dominated dwarf galaxies can be confidently distinguished from dark-matter-free star clusters, as well as to avoid low-mass subhalos where resolution effects become important in \SMW and \MWest.
However, both theoretical and observational arguments suggest that a significant population of satellite galaxies may exist with smaller sizes and lower luminosities (e.g., \citep{Manwadkar:2022, Cerny:2026}).
To explore the impact of such a satellite population, we remove the cuts on $M_V$ and $r_{1/2}$ from the ultimate scenario, while keeping the subhalo resolution cuts on $M_{\rm peak}$, $V_{\rm peak}$, and $V_{\rm max}$ (Sec.~\ref{sec:simulations}).

We validate our implementation against the results of \citet{LSSTDESC:2025hol} by generating discrete catalogs using the LSST selection function alone and counting detectable satellites with $M_V \lesssim 0$, $r_{1/2}>10~\mathrm{pc}$, and $D<300~\mathrm{kpc}$. For \SMW, $10^4$ realizations yield $N_{\mathrm{LSST}}=88\pm21$, in good agreement with the $89\pm20$ LSST-detectable Milky Way satellites reported in \citet{LSSTDESC:2025hol} for the idealized case of perfect star-galaxy separation. This agreement provides a consistency check of our implementation of the LSST observational selection function and discrete satellite detection procedure. 

Applying the same procedure to \MWest with $10^5$ realizations yields $N_{\mathrm{LSST}}=62\pm18$. Because we apply identical selection functions and analysis cuts to both suites, the lower \MWest prediction reflects differences in the underlying host-halo ensembles and satellite populations rather than differences in the LSST sensitivity.
We note that the galaxy--halo connection parameters used for both suites were calibrated using the two \SMW hosts. Independently calibrating the model using \MWest and the observed Milky Way satellite population could, in principle, bring the two predictions into closer agreement. We leave this recalibration to future work.

\begin{figure*}
    \centering
    \includegraphics[width=0.95\textwidth]{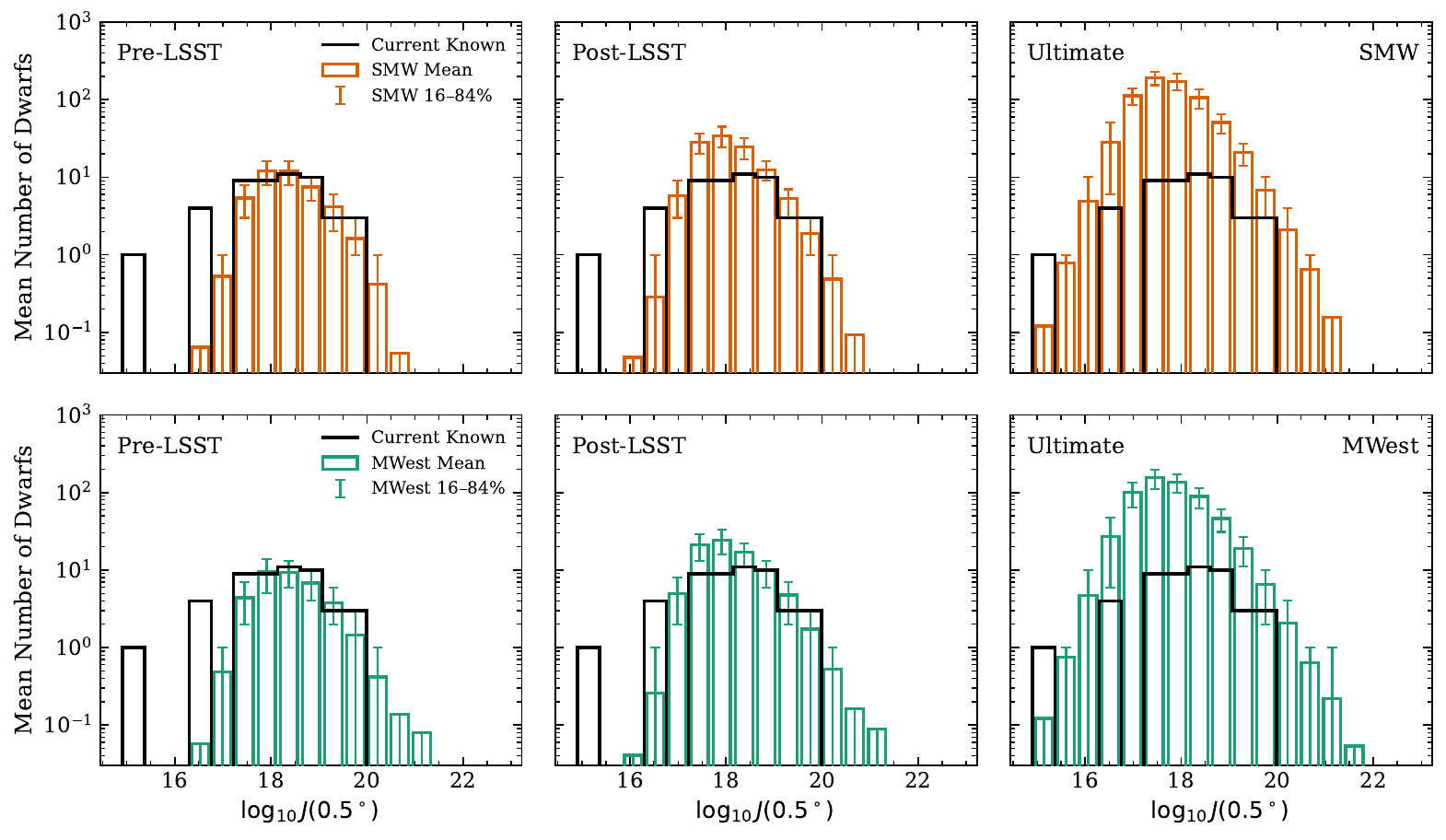}
    \caption{The $J$-factor distributions of simulated pre-LSST (left), post-LSST (middle) and ultimate (right) dwarf galaxy populations from the \SMW (top) and \MWest (bottom) simulation suites, compared with those of the currently known Milky Way satellites~\cite{McDaniel:2023bju}. Both suites broadly reproduce the observed pre-LSST distribution, particularly for the high $J$-factors that are most relevant for indirect dark matter searches.}
    \label{fig:j_factor}
\end{figure*}

\section{J-Factor Calculation and Forecast Methods}
\label{sec:MC}

Our pipeline, \BabyBison, simulates gamma-ray photon counts from dwarf galaxies and their backgrounds and performs a joint likelihood analysis to obtain a 95\% confidence level (CL) upper limit on $\langle\sigma v\rangle$ for each realization. The dark matter signal model, background treatment, and statistical analysis follow the approach described in Ref.~\cite{XuHooper2023}, as summarized below.

\subsection{Dark Matter Signal and Galaxy Modeling}
The gamma-ray flux from annihilating dark matter in a given dwarf galaxy can be written as
\begin{equation}
    \frac{d\Phi_\gamma}{dE_\gamma}
    \;=\; \frac{\langle\sigma v\rangle}{8\pi\,m_\chi^{2}}
    \left.\frac{dN_\gamma}{dE_\gamma}\right|_{\mathrm{ann}} J(\Delta\Omega),
    \label{eq:dm_signal}
\end{equation}
where $d\Phi_\gamma/dE_\gamma$ is the differential photon flux, $\langle\sigma v\rangle$ is the thermally averaged annihilation cross section, $m_\chi$ is the dark matter mass, $\left.dN_\gamma/dE_\gamma\right|_{\mathrm{ann}}$ is the per-annihilation photon spectrum for the chosen final state, and $J(\Delta\Omega)$ is the line-of-sight (los) integral of the square of the dark matter density over a solid angle, $\Delta\Omega$:
\begin{equation}
    J(\Delta\Omega)
    =
    \int_{\Delta\Omega}
    \int_{\mathrm{los}}
    \rho_{\mathrm{DM}}^2(\mathbf{r})\, ds\, d\Omega.
\end{equation}
We calculate the per-annihilation photon spectra using the publicly available PPPC4DMID package~\cite{Cirelli:2010xx,Ciafaloni:2010ti}, although \BabyBison also supports COSMIX~\cite{Arina:2023eic}.  

We assign each simulated galaxy a $J$-factor using the empirical scaling relation of Ref.~\cite{PaceStrigari2019}. This relation depends on the satellite luminosity $L$, heliocentric distance $D$, and half-light radius $r_{1/2}$:
\begin{align}
\label{eq:pace_strigari}
    \log_{10} &\bigg[\frac{J\left(0.5^\circ\right)}{\mathrm{GeV^2\,cm^{-5}}}\bigg]
    = 18.14
    + 0.23 \log_{10}\!\left(\frac{L}{10^6 \, L_\odot}\right) \\
    \nonumber
    &\quad - 2.0 \log_{10}\!\left(\frac{D}{100 \, \mathrm{kpc}}\right)
    - 0.5 \log_{10}\!\left(\frac{r_{1/2}}{300\,\mathrm{pc}}\right).
\end{align}
Here, $J(0.5^{\circ})$ denotes the $J$-factor integrated within an angular radius of $0.5^{\circ}$. At fixed luminosity, the relation assigns larger $J$-factors to nearby, compact systems. We draw $\log_{10}[J/(\mathrm{GeV^2\,cm^{-5}})]$ from a Gaussian distribution centered on the value predicted by Eq.~\ref{eq:pace_strigari}, and with a standard deviation of $\sigma_{\log_{10} J}=0.27$ \cite{PaceStrigari2019}. Note that, in our galaxy--halo connection model, the dwarf galaxy properties that enter Eq.~\ref{eq:pace_strigari} are set by global properties of each subhalo (e.g., maximum circular velocity) rather than their detailed density profiles. Note that Eq.~\ref{eq:pace_strigari} does not include any enhancements to the $J$-factor from prompt cusps~\cite{Delos:2022bhp}, making our predictions conservative in this regard.

In Fig.~\ref{fig:j_factor}, we compare the simulated $J$-factor distributions with those of the currently known satellites~\cite{McDaniel:2023bju}. The left, middle, and right columns show the pre-LSST, post-LSST, and ultimate populations, respectively, with SMW in the top row and MWest in the bottom row. The colored histograms show the mean number of dwarfs in each bin across 30{,}000 \SMW and \MWest realizations, with error bars spanning the central 68\% intervals. 

Both suites broadly reproduce the observed pre-LSST $J$-factor distribution, particularly at the high $J$-factors that are most important for dark matter annihilation searches. Although the observed $J$-factor distribution was not directly used to calibrate the galaxy--halo connection model, some level of agreement is expected because the quantities entering the empirical $J$-factor relation are themselves constrained by the observed satellite population, as is the galaxy--halo connection model.

The additional satellites detected by LSST largely occupy the same $J$-factor range as the pre-LSST sample, leaving the normalized distributions broadly similar in shape. The slight deficit of simulated systems at the lowest observed $J$-factors has little impact on the projected constraints because these systems contribute negligibly to the annihilation signal. 

In contrast, the ultimate sample contains appreciably more satellites with large $J$-factors. Quantitatively, the maximum $J$-factor at which the simulations predict one or more satellites per realization moves from $\log_{10} [J/(\mathrm{GeV^2\,cm^{-5}})] \sim 19.8$ in the post-LSST sample to $\log_{10} [J/(\mathrm{GeV^2\,cm^{-5}})] \sim 20.5$ in the ultimate sample. This shift is largely due to the fact that the ultimate sample removes the observational selection cuts on small, faint satellites (i.e., $M_V > 0$ and $r_{1/2} < 10\,{\rm pc}$), which may have evaded detection by current surveys even at relatively small distances and are explicitly removed from the post-LSST sample. Since the gamma-ray sensitivity is largely driven by a small number of dwarf galaxies with the highest $J$-factors, this difference has an appreciable effect on the sensitivity to dark matter annihilation with the ultimate sample. These results are consistent with  studies of recently discovered ultra-faint compact satellites \citep{Crnogorcevic:2023ijs, Circiello:2025}.

Throughout this analysis, we use $J$-factors integrated within $0.5^\circ$ and adopt a fixed-aperture photon-counting likelihood. The limitations of this approximation are most relevant for the nearest systems, which have the largest angular extensions.

\subsection{Background Modeling}

Our gamma-ray background model consists of unresolved point sources, an extragalactic isotropic component, and a position-dependent Galactic diffuse component (\texttt{gll\_iem\_v07.fits}). We model the unresolved point-source population as being blazar-dominated and describe their source-count distribution using the triply broken power law described in Ref.~\cite{Marcotulli:2020fpm}. We statistically marginalize over this population across the photon-flux range of $10^{-10} \leq S \leq 10^{-6}\ {\rm ph\,cm^{-2}\,s^{-1}}$. Each unresolved source is assigned the spectral shape of the extragalactic gamma-ray background (EGB), using the \mbox{Fermi-LAT} foreground
Model A~\cite{Ackermann:2014usa}. We obtain the residual isotropic component by subtracting the modeled unresolved point-source contribution from the total EGB.

Fluctuations in the number and fluxes of unresolved sources between lines of sight produce an intrinsically non-Poissonian photon-count distribution. We combine this distribution with the isotropic and Galactic diffuse contributions to construct the background-count probability distribution for each galaxy and energy bin, $P^\mathrm{bg}_{g,i}(N_{\mathrm{bg}})$, where $N_{\mathrm{bg}}$ denotes the
number of background photons, and $g$ and $i$ label the galaxy and energy bin, respectively.

\subsection{Synthetic Data, Likelihood, and Upper Limits}

\BabyBison partitions the simulated galaxy populations described in  Sec.~\ref{sec:simulations} into $\mathcal{O}(10^{3})$ batches, each representing an independent realization of a stacked dwarf galaxy analysis. Each batch contains $\mathcal{O}(10^{1})$--$\mathcal{O}(10^{3})$ dwarf galaxies and the number of batches is user-configurable. For each realization, we select an observer position and orient the satellite system relative to the LMC as described above and then apply the survey selection functions and Galactic-plane cut. We draw the dwarf galaxy $J$-factors, compute the position-dependent background distributions, and generate a synthetic gamma-ray data set. Holding the galaxy sample, model parameters, $J$-factors, and synthetic data each fixed, we evaluate the likelihood to obtain a 95\% CL upper limit on $\langle\sigma v\rangle$ in each realization and at each mass. We then determine the median 95\% CL upper limit on $\langle\sigma v\rangle$ as a function of $m_\chi$, together with the 68\% and 95\% containment intervals across all realizations.

We use 16 logarithmically-spaced energy bins, with different energy ranges for the validation and forecasts, as specified below. Both the energy binning and instrumental response are configurable. The code can apply a uniform exposure in each energy bin, accounting for the telescope's energy-dependent effective area. Alternatively, Fermi-LAT spacecraft files can be used with the instrument response to calculate a sky-dependent exposure over the selected energy range and observing intervals.

To forecast annihilation cross section upper limits, we generate background-only photon counts, $N^{\mathrm{obs}}_{g,i}$, by drawing from $P^\mathrm{bg}_{g,i}$ for each galaxy and energy bin, corresponding to the case of the null hypothesis, $\langle\sigma v\rangle = 0$. At a given dark matter mass, the joint log-likelihood is given by
\begin{equation}
    \ln \mathcal{L}(\langle\sigma v\rangle)
    \;=\; \sum_{g }\, \sum_{i}
    \ln P^\mathrm{tot}_{g,i}\!\left(N^{\mathrm{obs}}_{g,i} \,\big|\,
    \langle\sigma v\rangle, m_\chi, J_g\right),
    \label{eq:logL}
\end{equation}
where $P^\mathrm{tot}_{g,i}$ is the signal-plus-background photon-count probability distribution. We define the test statistic as
\begin{equation}
    \mathrm{TS}_{\langle\sigma v\rangle}
= -2\left[\ln\mathcal{L}\left(\langle\sigma v\rangle\right)
- \ln\mathcal{L}\left(\widehat{\langle\sigma v\rangle}\right)\right],
    \label{eq:TS}
\end{equation}
where $\widehat{\langle\sigma v\rangle}$ is the best-fit cross section at a fixed value of $m_\chi$, restricted to the physical range $\langle\sigma v\rangle \geq 0$.  For each realization and mass, we obtain the one-sided 95\% CL upper limit by finding the cross section above the best-fit value at which $\mathrm{TS}_{\langle\sigma v\rangle}=2.71$.

\subsection{Parameter Estimation}
\label{sec:gce2d}

In addition to forecasting upper limits on the annihilation cross section, \BabyBison can be used to predict the statistical significance at which a dark matter signal could be established, and to evaluate the precision with which its parameters could be measured. As an example, we consider a scenario motivated by dark matter interpretations of the Galactic Center gamma-ray excess (GCE)~\cite{Goodenough:2009gk,Hooper:2010mq,Hooper:2011ti,Abazajian:2012pn,Hooper:2013rwa,Gordon:2013vta,Daylan:2014rsa,Calore:2014xka,Fermi-LAT:2015sau,DiMauro:2026fnp,Cholis:2021rpp,DiMauro:2021raz}, featuring a dark matter particle with a mass of $40\,\mathrm{GeV}$ and an annihilation cross section of $\langle\sigma v\rangle = 2\times 10^{-26}\,\mathrm{cm^{3}\,s^{-1}}$ to $b\bar{b}$. We inject this signal into the simulated data for each realization and maximize the joint log-likelihood in Eq.~\eqref{eq:logL} over both parameters to obtain the best-fit values, $\hat{m}_\chi$ and $\widehat{\langle\sigma v\rangle}$. To quantify the preference for a signal over the background-only hypothesis, we define the following test statistic:
\begin{equation}
    \mathrm{TS}_{0}
    \equiv -2\left[\ln\mathcal{L}\left(\langle\sigma v\rangle=0\right)-\ln\mathcal{L}\left(\hat{m}_\chi, \widehat{\langle\sigma v\rangle}\right)\right].
\end{equation}
We perform this evaluation for the pre-LSST dwarf galaxy population, the post-LSST galaxy population, and the complete modeled Milky Way dwarf galaxy population.

If the dark matter mass were fixed by an independent measurement (such as that of the gamma-ray emission from the Galactic Center), we could perform a single-parameter fit to $\langle\sigma v\rangle$. In the asymptotic (large-count) limit, $\mathrm{TS}_0$ is distributed as $\tfrac{1}{2}\chi^2_0 + \tfrac{1}{2}\chi^2_1$, and the one-sided Gaussian significance is $\sqrt{\mathrm{TS}_{0}}$ \citep{Wilks:1938,Chernoff:1954}. However, when scanning over both $\langle\sigma v\rangle$ and $m_\chi$, the asymptotic results do not apply because $m_\chi$ is not identifiable under the null hypothesis (i.e., $m_\chi$ drops out when $\langle \sigma v \rangle = 0$) \citep{Davies:1987}.
In this case, converting $\mathrm{TS}_0$ into a global detection significance requires the calculation of a look-elsewhere correction \citep{Gross:2010}. 
Moreover, in real data the distribution of $\mathrm{TS}_0$ can depart from the asymptotic expectation due to background mis-modeling, unresolved sources, etc. 
In this case, the null distribution of $\mathrm{TS}_0$ can be calibrated empirically on blank-sky regions, which accounts for both look-elsewhere and mis-modeling effects \citep{Ackermann:2014, FermiDwarfs2015}. 
Due to these complications, we do not attempt to report a global significance in our analysis and instead focus on measurement precision.

\begin{figure*}
    \centering
    \includegraphics[width=\textwidth]{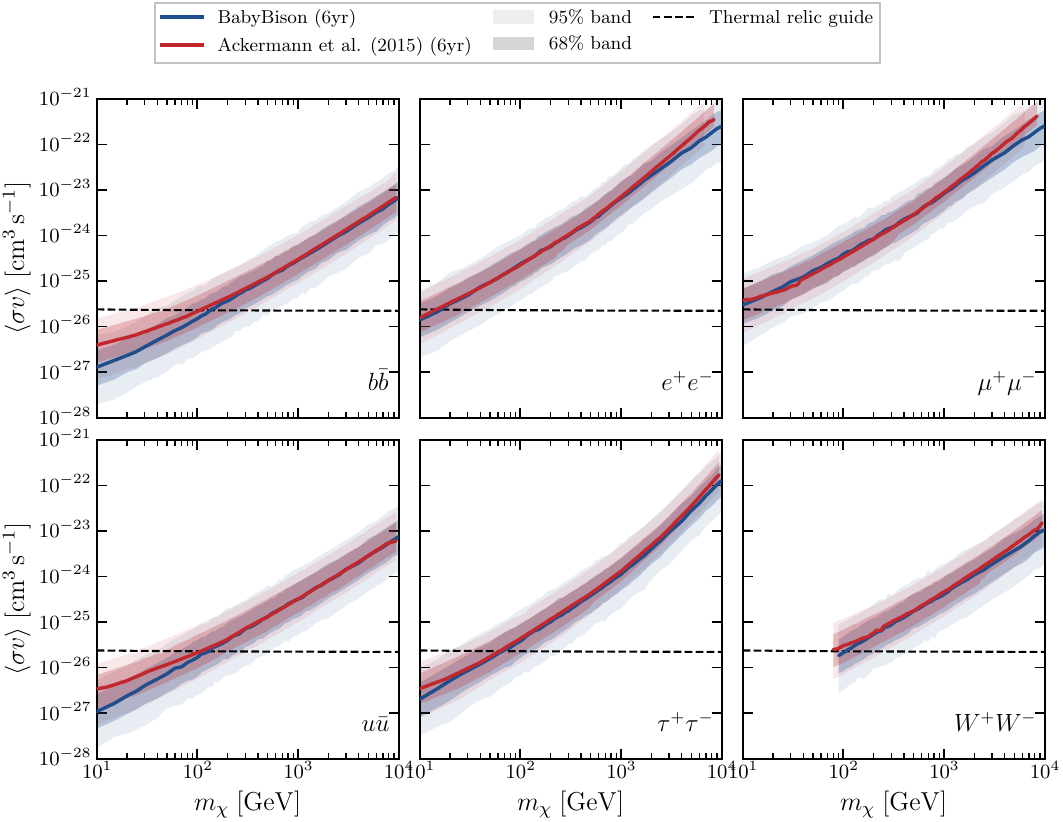}
    \caption{A comparison of the projected sensitivity of Fermi-LAT to dark matter annihilation in dwarf galaxies obtained using the \BabyBison pipeline compared with the results published by the Fermi-LAT Collaboration~\cite{FermiDwarfs2015}. In each panel, the blue curve denotes the median 95\% CL upper limit on $\langle\sigma v\rangle$ obtained using \BabyBison, while the red curve is the constraint from Fermi. These curves are each surrounded by the corresponding 68\% and 95\% containment intervals. The dashed horizontal line represents the (velocity-independent) cross section required of a dark matter candidate that is a thermal relic of the early universe.}
    \label{fig:allchan_vsfermi6yr}
\end{figure*}

To forecast the precision of parameter measurements, we evaluate
\begin{equation}
   \mathrm{TS}_{m_\chi,\langle\sigma v\rangle} \equiv -2\left[\ln\mathcal{L}\left(m_\chi, \langle\sigma v\rangle\right)-\ln\mathcal{L}\left(\hat{m}_\chi, \widehat{\langle\sigma v\rangle}\right)\right],
\end{equation} 
at each point in the  $(m_\chi, \langle\sigma v\rangle)$ plane for each realization. Under the usual asymptotic likelihood-ratio approximation, thresholds of 2.30 and 6.18 define the joint 68.3\% and 95.4\% confidence regions for the two parameters in an individual realization. We then take the median value of $\mathrm{TS}_{m_\chi,\langle\sigma v\rangle}$ at each point in that plane and draw $1\sigma$ and $2\sigma$ confidence contours around the median best-fit point.

\section{Results}
\label{sec:results}

In most of the figures shown in this section, the central curve represents the median result obtained across all of the realizations in the ensemble, while the inner and outer shaded bands enclose the 68\% and 95\% containment intervals. These bands reflect the variation among the simulated satellite populations, the uncertainty in the galaxy--halo connection model, and the scatter in the assigned $J$-factors, as well as statistical fluctuations in the simulated photon counts.

\subsection{Pipeline Validation and Comparison with Fermi-LAT Limits}
\label{sec:res_validation}

Before applying \BabyBison to the simulated satellite populations introduced in Sec.~\ref{sec:simulations}, we validate the signal, background, and likelihood implementation described in Sec.~\ref{sec:MC} against the Fermi-LAT Collaboration's six-year dwarf galaxy analysis~\cite{FermiDwarfs2015}. We use the same catalog of kinematically confirmed Milky Way dwarf galaxies, $J$-factors, and exposure window as Ref.~\cite{FermiDwarfs2015} to produce upper limits on the dark matter annihilation cross section.

\begin{figure*}
    \centering
    \includegraphics[width=0.95\textwidth]{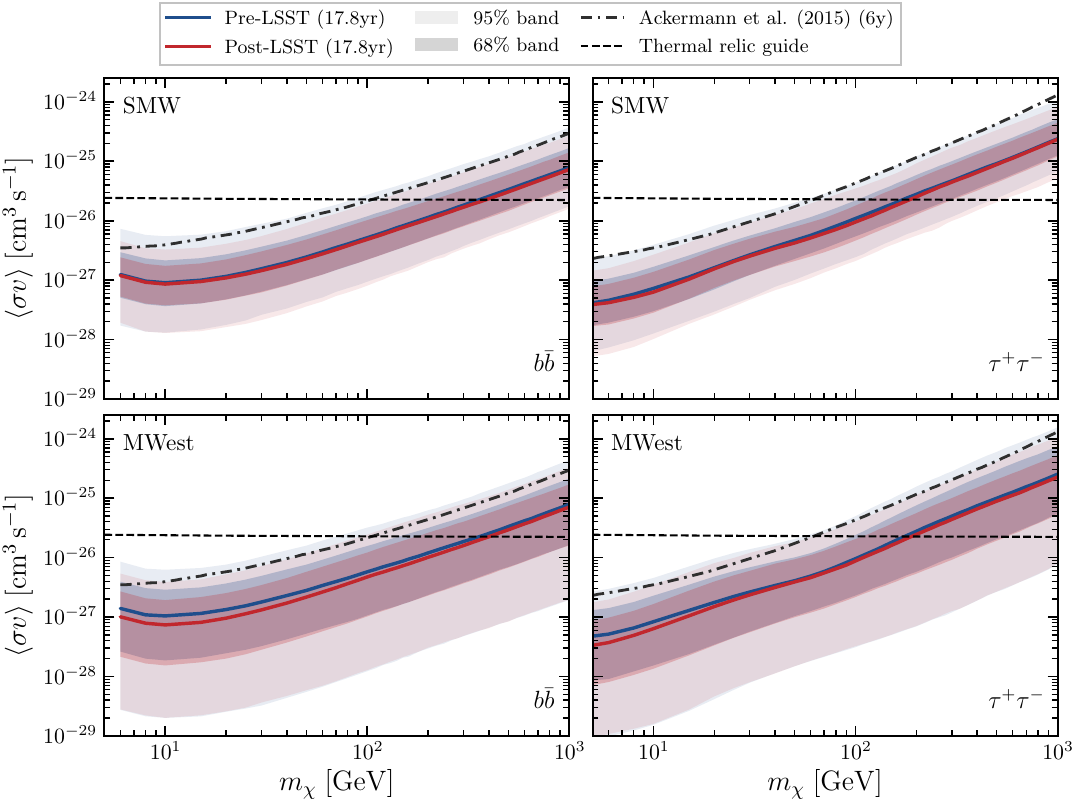}
    \caption{The distribution of 95\% CL upper limits on $\langle \sigma v\rangle$ for dark matter annihilation into $b\bar{b}$ (left) and $\tau^+\tau^-$ (right), using the pre-LSST (blue) and post-LSST (red) satellite catalogs drawn from the same realizations described in Sec.~\ref{sec:simulations}. The top and bottom rows show the \SMW and \MilkyWayest suites, respectively. Solid curves denote the median limit at each mass, while the darker and lighter shaded regions enclose the central 68\% and 95\% intervals across realizations, respectively. Both suites predict only modest improvements from the additional LSST satellites. The dashed black curves mark the benchmark thermal relic cross section. These forecasts assume 17.8 years of exposure. For comparison, the dash-dotted curve denotes the published median expected limit from the six-year Fermi-LAT analysis~\cite{FermiDwarfs2015}.}
    \label{fig:run2_before_after}
\end{figure*}

In Fig.~\ref{fig:allchan_vsfermi6yr}, we compare our validation results to the limits published by the Fermi-LAT Collaboration, for six representative annihilation channels~\cite{FermiDwarfs2015}. In each case, the sensitivity bands obtained with \BabyBison agree reasonably well with the Fermi-LAT results. The largest differences occur at low dark matter masses, where the comparison is particularly sensitive to the energy threshold. Our analysis uses photons between $1\,\mathrm{GeV}$ and $1\,\mathrm{TeV}$, whereas the Fermi-LAT analysis uses photons between $500\,\mathrm{MeV}$ and $500\,\mathrm{GeV}$. For low-mass dark matter, a larger fraction of the annihilation spectrum lies below $1\,\mathrm{GeV}$, such that this difference in energy threshold changes the signal and background contributions to the likelihood. This likely contributes to the discrepancy observed at the lowest masses, although differences in the background treatment and spatial likelihood may also play a role.

\begin{figure*}
    \centering
    \includegraphics[width=0.95\textwidth]{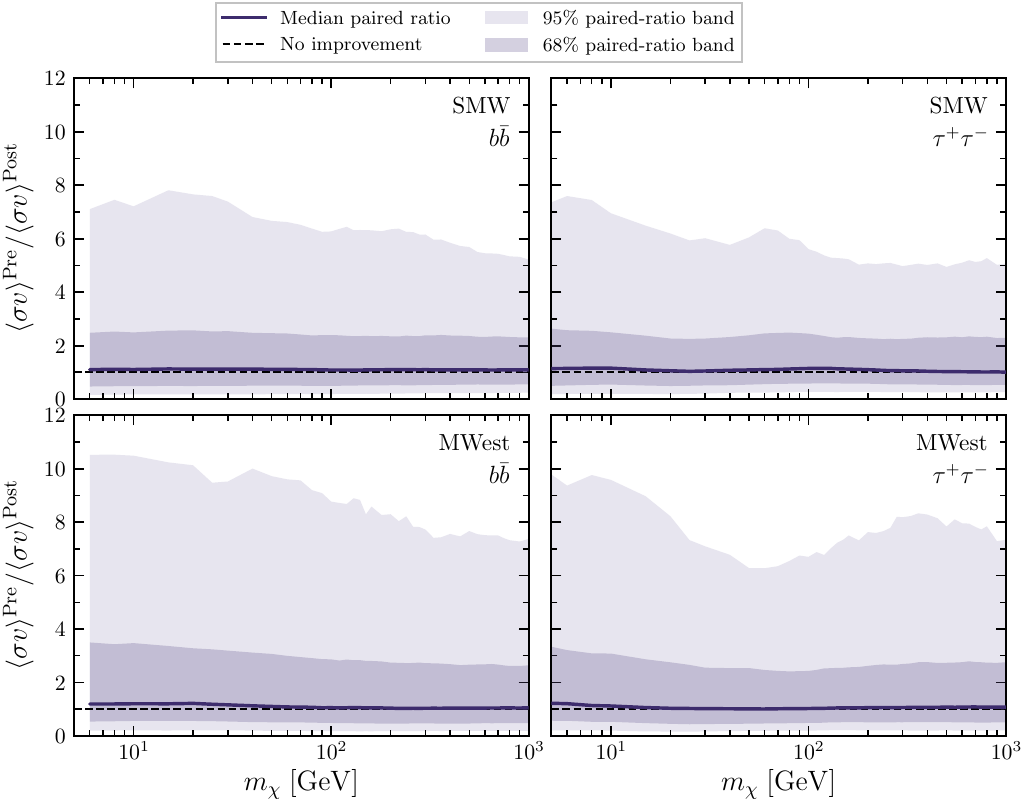}
    \caption{The distribution of the LSST improvement ratio, $R(m_\chi)=\langle\sigma v\rangle^{\mathrm{pre}}/ \langle\sigma v\rangle^{\mathrm{post}}$, for dark matter annihilation into $b\bar{b}$ (left) and $\tau^+\tau^-$ (right), using the \SMW (top) and \MWest (bottom) simulation suites. The ratio compares the pre-LSST and post-LSST 95\% CL upper limits within each realization, with $R>1$ indicating improved sensitivity. Solid purple curves denote the median paired ratio at each mass, while the darker and lighter shaded regions enclose the central 68\% and 95\% intervals across realizations, respectively. Both suites predict modest improvements in typical realizations, although much more substantial gains can be obtained in relatively rare cases.}
    \label{fig:lsst_ratio}
\end{figure*}

\subsection{Projected LSST Constraints}
\label{sec:res_lsst}

Having validated \BabyBison against the published benchmark from the Fermi-LAT Collaboration, we now assess how future LSST dwarf galaxy discoveries could improve the sensitivity of Fermi-LAT searches for annihilating dark matter. We analyze an ensemble of 2{,}000 realizations using gamma rays with energies between $1\,\mathrm{GeV}$ and $1\,\mathrm{TeV}$ in $16$ logarithmically-spaced bins. In Fig.~\ref{fig:run2_before_after}, we show the resulting 95\% CL upper limits for $\chi\chi\to b\bar{b}$ (left) and $\chi\chi\to\tau^{+}\tau^{-}$ (right), using the pre-LSST (blue) and post-LSST (red) dwarf galaxy populations from the \SMW (top) and \MWest (bottom) simulation suites. Within each realization, we hold the underlying satellite population and Monte Carlo inputs fixed, ensuring that the differences between the results in each panel are due entirely to the different dwarf galaxy populations being considered. The median pre-LSST and post-LSST results are similar in both suites, demonstrating the modest impact of anticipated LSST dwarf discoveries on the median expectation for such an analysis.  
Quantitatively, the median \SMW pre-LSST limits intersect the thermal-relic benchmark at $m_\chi\simeq350\,\mathrm{GeV}$ for $b\bar{b}$ and $170\,\mathrm{GeV}$ for $\tau^{+}\tau^{-}$. After LSST, these crossings shift to $380\,\mathrm{GeV}$ and $180\,\mathrm{GeV}$, respectively. The corresponding crossings for \MWest shift from $350\,\mathrm{GeV}$ to $400\,\mathrm{GeV}$ for $b\bar{b}$ and from $170\,\mathrm{GeV}$ to $190\,\mathrm{GeV}$ for $\tau^{+}\tau^{-}$.

These gains are smaller than might be expected from the increase in the number of targets. If all dwarfs carried comparable statistical weight in the background-dominated regime, the upper limit would scale approximately as $N_{\rm dwarf}^{-1/2}$. The increase from approximately 42 to 104 detectable satellites in \SMW, and from 35 to 81 in \MWest, would then be expected to improve the limits by factors of $\sqrt{104/42}\simeq1.57$ and $\sqrt{81/35}\simeq1.52$, respectively. The smaller gains in the full likelihood analysis indicate that the newly detected satellites carry less statistical weight, on average, than the pre-LSST targets. This is consistent with the joint sensitivity being dominated by a small number of systems with large $J$-factors.

To quantify the improvement in the projected constraints from future LSST dwarf-galaxy discoveries, we define
\begin{equation}
R(m_\chi) \equiv
\frac{\langle\sigma v\rangle^{\mathrm{Pre}}}
     {\langle\sigma v\rangle^{\mathrm{Post}}},
\end{equation}
where $\langle\sigma v\rangle^{\mathrm{Pre}}$ and
$\langle\sigma v\rangle^{\mathrm{Post}}$ are the pre-LSST and post-LSST 95\% CL upper limits at the same mass. Thus, $R>1$ indicates an improvement in sensitivity, and $1-1/R$ gives the fractional reduction in the upper limit.

In Fig.~\ref{fig:lsst_ratio}, we show the distribution of $R$ for the \bbbar (left) and \tptm (right) channels, using the \SMW (top) and \MWest (bottom) simulation suites. The solid purple curve denotes the median paired ratio at each mass, and the shaded regions enclose the central 68\% and 95\% containment intervals.

Across the mass range shown, the \SMW median improvement ratio ranges from approximately 1.09 to 1.15 for \bbbar and from 1.03 to 1.19 for \tptm. This corresponds to reductions in the upper limits of approximately 8.1--12.9\% and 2.7--15.7\%, respectively. The upper boundary of the central 68\% interval ranges from $R\simeq2.37$ to 2.71 for \bbbar and from 2.32 to 2.79 for \tptm, while the upper boundary of the central 95\% interval reaches $R\sim8.36$--$8.44$. Thus, although the typical improvement is modest, some realizations yield substantially larger gains.

For \MWest, the median ratio ranges from $R\simeq1.03$ to $1.22$ for \bbbar and from $R\simeq1.01$ to $1.22$ for \tptm, corresponding to reductions in the upper limits of $2.9\%$--$18.0\%$ and $0.6\%$--$18.1\%$, respectively. The upper boundary of the central 68\% interval ranges from $R\simeq2.71$ to $3.32$ for \bbbar and from $R\simeq2.52$ to $3.26$ for \tptm, broadly comparable to the \SMW results.

\begin{figure*}
    \centering
    \includegraphics[width=\textwidth]{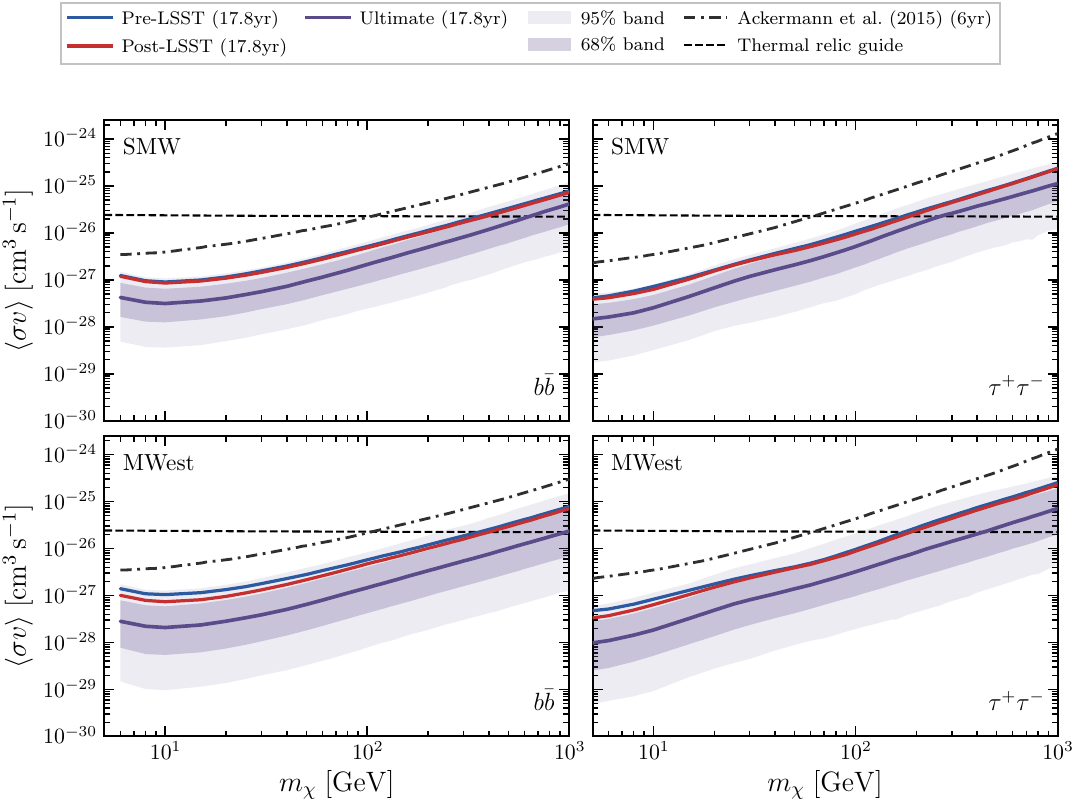}
    \caption{Projected 95\% CL upper limits on $\langle \sigma v \rangle$ for dark matter annihilation into $b\bar{b}$ (left) and $\tau^{+}\tau^{-}$ (right), using the ultimate catalog from the \SMW (top) and \MWest (bottom) simulation suites and assuming 17.8 years of Fermi-LAT data. Each ensemble contains 2{,}000 realizations, totaling $\mathcal{O}(10^{6})$ simulated galaxies. Solid curves denote the median limit at each mass, while the darker and lighter shaded regions enclose the central 68\% and 95\% intervals across realizations, respectively. The dashed black curves mark the thermal-relic benchmark, and the dash-dotted curves show the six-year limit from Fermi-LAT~\cite{FermiDwarfs2015}. The \MWest median upper limits are lower than those of \SMW by median factors of approximately $1.5$ for $b\bar{b}$ and $1.7$ for $\tau^+\tau^-$, although the containment bands overlap substantially.}
    \label{fig:ult_fermi}
\end{figure*}

\begin{figure*}
    \centering
    \includegraphics[width=\textwidth]{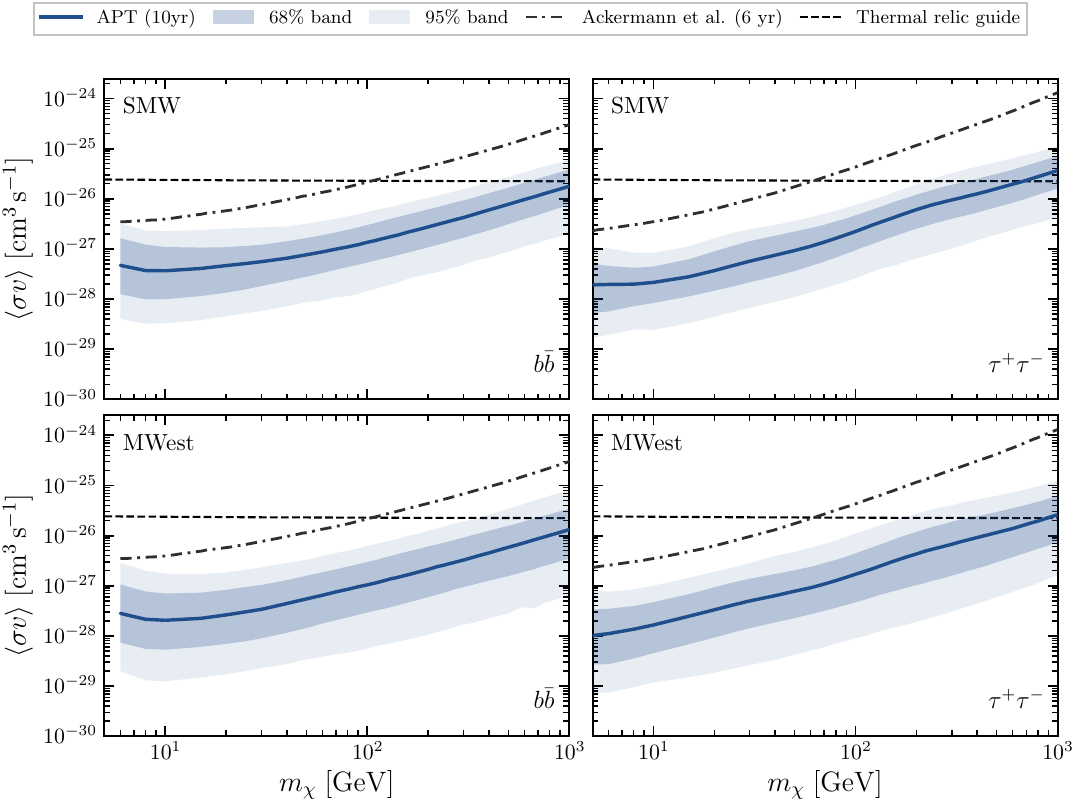}
    \caption{Similar to Fig.~\ref{fig:ult_fermi}, but considering 10 years of data from the proposed Advanced Particle-astrophysics Telescope (APT) \citep{APT:2021lhj}.}
    \label{fig:ult_apt}
\end{figure*}

\begin{figure}[t]
    \centering
    \includegraphics[width=\columnwidth]
    {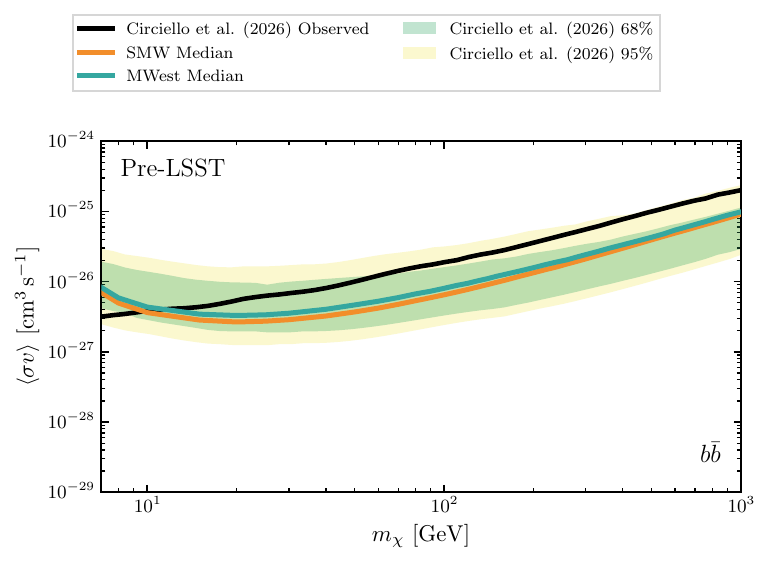}
    \caption{
    A comparison of the expected 95\% CL upper limit on $\langle \sigma v \rangle$ for dark matter annihilation into $b\bar{b}$ from the pre-LSST satellite populations in the Symphony Milky Way (SMW) and Milky Way-est (MWest) simulations with those from the Fermi-LAT Legacy analysis of Circiello et al.~\cite{Circiello:2026dSph}. All analyses use 15.9 years of Fermi-LAT observations and a photon energy range of 2 to 1000 GeV. The orange and cyan curves represent the median expected limits from 2{,}000 realizations drawn from the SMW and MWest simulations, respectively. The black curve shows the observed constraint from Circiello et al., while the green and yellow shaded regions indicate the corresponding 68\% and 95\% containment intervals for the expected limit based on a blank fields (background only) analysis. 
    }

    \label{fig:fermi_prelsst_MWEST_SMW}
\end{figure}

\subsection{Sensitivity to the Ultimate Dwarf Catalog}
\label{sec:res_ult}

We define the \emph{ultimate dwarf catalog} as the complete population of Milky Way satellite galaxies produced by the galaxy--halo model outside the Galactic-plane mask, $|b|<10^{\circ}$. This idealized catalog provides a benchmark for the sensitivity that could be attained with a complete satellite census. 

We analyze 2{,}000 realizations from each of the \SMW and \MWest suites for the ultimate dwarf catalog, yielding $\mathcal{O}(10^{6})$ simulated galaxies per ensemble. We forecast limits for Fermi-LAT with 17.8 years of data and for the proposed Advanced Particle-astrophysics Telescope (APT) with 10 years of data~\cite{Buckley:2021LE}. APT is designed to provide a substantially larger effective area than Fermi-LAT while maintaining comparable angular resolution~\cite{Chen:2021APT}. Both forecasts use photons between $1\,\mathrm{GeV}$ and $1\,\mathrm{TeV}$ in 16 logarithmically-spaced energy bins. 

In Fig.~\ref{fig:ult_fermi}, we show the projected Fermi-LAT limits for the ultimate catalogs, with \SMW and \MWest in the top and bottom frames, and for annihilations into $b\bar{b}$ and $\tau^{+}\tau^{-}$ in the left and right frames, respectively. The dash-dotted curves show the published median expected limits from the six-year Fermi-LAT analysis~\citep{FermiDwarfs2015}. For \SMW, the median $b\bar{b}$ limit is roughly an order of magnitude below this benchmark and intersects the thermal-relic cross section at $m_\chi \simeq 620\,\mathrm{GeV}$. The corresponding crossing for $\tau^{+}\tau^{-}$ occurs at $m_\chi \simeq 260\,\mathrm{GeV}$. Across the overlapping mass range, the median improvement factors relative to the six-year Fermi-LAT limits are approximately $10.1$ for $b\bar{b}$ and $8.6$ for $\tau^{+}\tau^{-}$.

The \MWest median limits are approximately a factor of $1.50$ stronger than those of \SMW for $b\bar{b}$ and $1.66$ stronger for $\tau^+\tau^-$. The \MWest containment bands are also wider. Although the two suites have similar normalized satellite $J$-factor distributions, their stacked sensitivities depend on the total number of satellites and on the number of high-$J$ systems in each realization. Differences in these quantities can therefore produce different limits even when the normalized distributions are similar.

Fig.~\ref{fig:ult_apt} shows the corresponding APT forecasts. At $m_\chi \lesssim 100\,\mathrm{GeV}$, the APT and Fermi-LAT sensitivities are comparable, with only modest improvements from APT. The advantage of APT increases toward higher masses, where its larger effective area becomes increasingly important. At $m_\chi=1\,\mathrm{TeV}$, the median APT limits for $b\bar{b}$ are stronger than the corresponding Fermi-LAT limits by factors of approximately 2.3 for \SMW and 1.8 for \MWest. For $\tau^{+}\tau^{-}$, the improvement factors are approximately 3.1 and 2.7, respectively.

The projected constraints shown in Figs.~\ref{fig:run2_before_after} and~\ref{fig:ult_fermi} are likely to be more stringent than those ultimately realized by Fermi-LAT, given the modest gamma-ray excesses already observed toward several individual dwarf galaxies, including Reticulum II, Sculptor, and Tucana II~\citep{DiMauro2022,McDaniel:2023bju}. These excesses have resulted in constraints on dark matter annihilation that are somewhat weaker than those expected in a typical background-only realization.

In Fig.~\ref{fig:fermi_prelsst_MWEST_SMW}, we compare the upper limits projected for our simulated pre-LSST satellite populations with the observed upper limit published in an updated, legacy analysis of Fermi-LAT data for the Milky Way's dwarf population using 15.9 years of data and photons with energies between 2 and 1000 GeV~\cite{Circiello:2026dSph}. Our projected constraints for both simulation suites lie near the median expected limit reported by Ref.~\cite{Circiello:2026dSph} in their background only analysis, whereas their observed limit is weaker by a factor of $\sim 2$ over most of the mass range considered. If these excesses arise from statistical fluctuations, we would expect the APT sensitivity to be consistent with the projections presented here. If they instead reflect genuine gamma-ray emission from these targets, APT would produce weaker limits than we have projected here, but could capably characterize the properties of the observed emission.

\subsection{Dependence of Annihilation Limits on Host Halo Mass}

More massive host halos are expected to contain a larger number of detectable satellite galaxies~\cite{Engler:2021}, potentially yielding stronger limits on dark matter annihilation. Across the \MWest sample, host halo mass is strongly correlated with the predicted number of detectable satellites, featuring a Spearman coefficient of $\rho_S = 0.791$ ($p = 3.33\times10^{-5}$) and a Pearson coefficient of $r_P = 0.820$ ($p = 9.42\times10^{-6}$). However, a larger satellite population will not necessarily contain more high-$J$ systems (for example, due to increased tidal disruption of nearby subhalos). Indeed, host halo mass is not significantly correlated with the summed $J$-factors of the three satellites with the largest $J$-factors, with $\rho_S=-0.050$ ($p=0.835$) and $r_P=-0.066$ ($p=0.781$). The number of detected satellites is likewise not significantly correlated with this sum, with $\rho_S=-0.050$ ($p=0.835$) and $r_P=-0.113$ ($p=0.635$).

These results are consistent with the strong distance dependence of the empirical $J$-factor relation used in our model, $J \propto D^{-2}$~\cite{PaceStrigari2019}. The stacked sensitivity therefore depends strongly on whether a realization contains a small number of nearby, high-$J$ objects. To examine this dependence, we define the $J$-weighted distance of the three highest-$J$ satellites in each realization as
\begin{equation}
D_J=\frac{\sum_{i=1}^{3} J_i D_i}{\sum_{i=1}^{3} J_i}.
\end{equation}
We then take the median result from the realizations for each host. Across the 20 hosts, this statistic is strongly anti-correlated with the summed $J$-factor of these satellites, with $\rho_S=-0.946$ ($p=3.11\times10^{-10}$) and $r_P=-0.887$ ($p=1.87\times10^{-7}$). This association is partly built into the adopted $J$-factor relation and the definition of $D_J$. Taken together, this suggests that host halo mass primarily tracks the overall satellite abundance, whereas annihilation sensitivity depends more strongly on the distances and properties of the highest-$J$ satellites.

One limitation is that our baryonic disruption prescription does not explicitly scale the strength of disk-induced disruption with host halo or disk mass. We apply the same prescription to all hosts, although individual subhalo disruption probabilities vary. If less massive hosts contain less massive disks, they may experience less baryonic disruption. Accounting for this dependence could alter the relation between host mass and surviving satellite abundance and represents a systematic uncertainty that is not included in this analysis.

\begin{figure}
    \centering
    \includegraphics[width=0.48\textwidth]{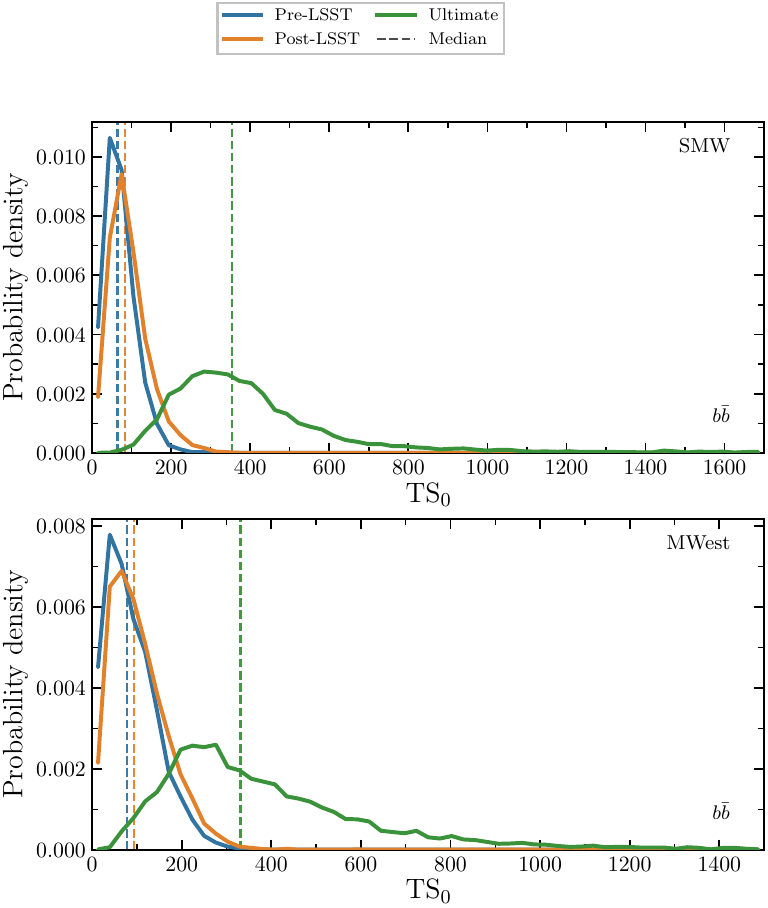}
    \caption{
    Distributions of the test statistic ($\mathrm{TS}_0$) for rejecting the null hypothesis when we inject a dark matter signal motivated by the observed characteristics of the Galactic Center gamma-ray excess, $m_\chi=40\,\mathrm{GeV}$ and  $\langle\sigma v\rangle=2\times10^{-26}\,\mathrm{cm^{3}\,s^{-1}}$ into $b\bar{b}$. Results are shown for the pre-LSST (blue), post-LSST (orange), and ultimate dwarf (green) catalogs, and for the \SMW (top) and \MWest (bottom) simulation suites. The vertical lines denote the median of each distribution.
    }
    \label{fig:mwest_injected_ts}
\end{figure}

\subsection{Injected-Signal Detectability}
\label{sec:res_ts}

The results presented in Sec.~\ref{sec:res_lsst} quantify the improvement in the expected upper limits from LSST discoveries under the background-only hypothesis. We now consider the complementary question of how significantly a dwarf galaxy analysis could detect a dark matter signal and how precisely it could recover the signal parameters. To address this question, we inject a benchmark signal motivated by dark matter interpretations of the Galactic Center gamma-ray excess (GCE), with $m_\chi=40\,\mathrm{GeV}$ and  $\langle\sigma v\rangle=2\times10^{-26}\,\mathrm{cm^{3}\,s^{-1}}$ into $b\bar{b}$. Using the test statistics defined in Sec.~\ref{sec:gce2d}, we evaluate detection significance and parameter recovery for the pre-LSST, post-LSST, and ultimate dwarf catalogs.

\subsubsection{Null Hypothesis Rejection}

In Fig.~\ref{fig:mwest_injected_ts}, we show the $\mathrm{TS}_0$ distributions for the pre-LSST (blue solid), post-LSST (orange solid), and ultimate dwarf (green solid) catalogs. The top and bottom panels show results for \SMW and \MWest, respectively. Vertical lines with matching colors and line styles denote the medians of the corresponding distributions. For both simulation suites, the distributions are strongly right-skewed, displaying long tails toward large values of $\mathrm{TS}_0$, particularly in the case of the ultimate dwarf catalog.  Again, however, it should be noted that unmodeled systematic uncertainties on the corresponding $J$-factors should be expected to somewhat reduce this sensitivity~\cite{Bertolez-Martinez:2026yqt}.

\subsubsection{Galactic Center Excess Parameter Reconstruction}
\label{sec:res_gce}

\begin{figure}
    \centering
    \includegraphics[width=\columnwidth]{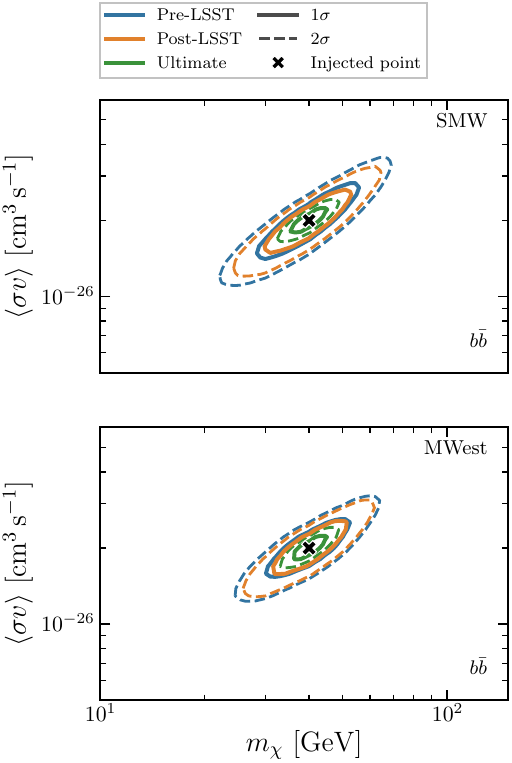}
    \caption{The parameter-recovery forecasts from 2{,}000 realizations of \SMW (top) and \MWest (bottom), assuming 17.8 years of Fermi-LAT data and injecting a dark matter signal motivated by the observed characteristics of the Galactic Center gamma-ray excess, $m_\chi=40\,\mathrm{GeV}$ and $\langle\sigma v\rangle=2\times10^{-26}\,\mathrm{cm^{3}\,s^{-1}}$ into $b\bar{b}$. Results are shown for the pre-LSST (blue), post-LSST (orange), and ultimate (green) catalogs. The solid and dashed contours represent the $1\sigma$ and $2\sigma$ forecast regions, respectively.
    }
\label{fig:gce_bb}
\end{figure}

In Fig.~\ref{fig:gce_bb}, we show the parameter-recovery forecasts from 2{,}000 realizations of the \SMW (top) and \MWest (bottom) simulation suites, assuming 17.8 years of Fermi-LAT data with the GCE benchmark signal injected. For each realization, we subtract the maximum log-likelihood from the log-likelihood surface and then take the pointwise median across realizations and shift the resulting surface so that its maximum is zero. Results are shown for the pre-LSST (blue), post-LSST (orange), and ultimate (green) catalogs. The solid and dashed contours correspond to $-2\Delta\ln\mathcal{L}=2.30$ and 6.18, respectively, and thus represent the $1\sigma$ and $2\sigma$ forecast regions.

For each simulation suite and catalog, the maximum of the combined surface coincides quite well with the injected benchmark parameters. As the catalog is expanded to include larger samples of satellite galaxies, the ability of Fermi-LAT to constrain these parameters increases. These results assume that unmodeled systematic uncertainties on the $J$-factors are not significant. Such uncertainties could appreciably degrade the ability of gamma-ray telescopes to accurately reconstruct the annihilation cross section of the dark matter particle~\cite{Bertolez-Martinez:2026yqt}.

\begin{figure*}
    \centering
    \includegraphics[width=\textwidth]
    {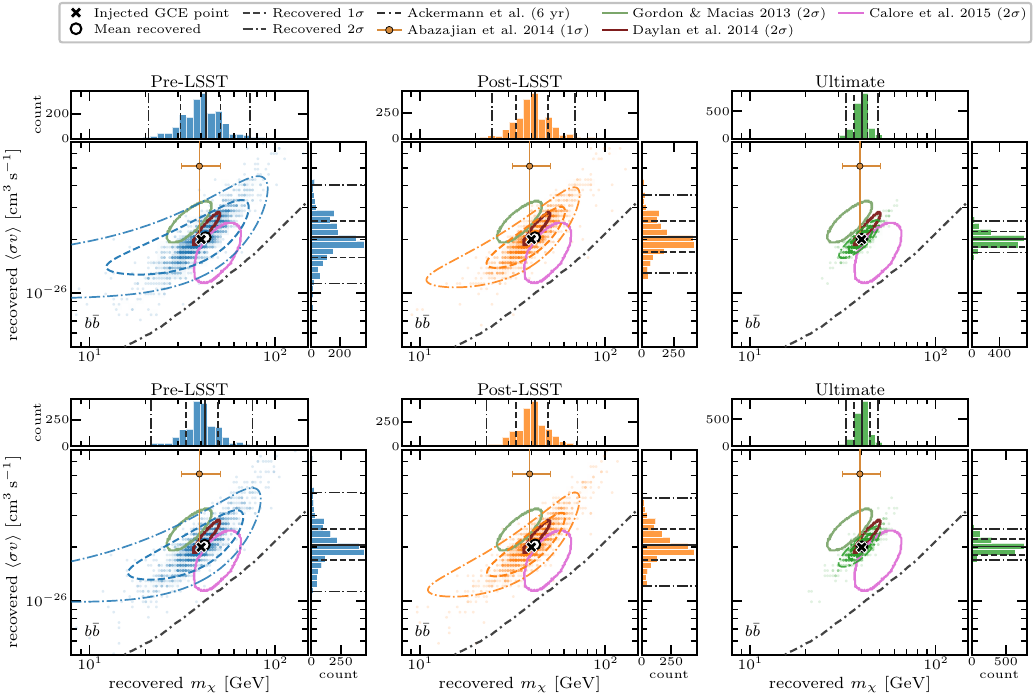}
    \caption{
    The distribution of recovered best-fit points across $N=2{,}000$ realizations for \SMW (top row) and \MWest (bottom row), shown for the pre-LSST (left), post-LSST (center), and ultimate dwarf (right) catalogs. For each scenario, the central panel shows the density of recovered best-fit points in the $(m_\chi, \langle\sigma v\rangle)$ plane, together with the injected point (black cross), the ensemble mean (open circle), and the $1\sigma$ (dashed) and $2\sigma$ (dot-dashed) contours. The flanking panels show the marginal distributions of the recovered best-fit values of $m_\chi$ and $\langle\sigma v\rangle$, with the ensemble mean indicated by the solid line and the corresponding $1\sigma$ (dashed) and $2\sigma$ (dot-dashed) intervals. Several previous analyses of the dark matter interpretation of the Galactic Center gamma-ray excess are also shown for reference
    \cite{Abazajian2014, GordonMacias2013, Daylan2014, Calore2015}.
    }
    \label{fig:gce_recovered}
\end{figure*}

Fig.~\ref{fig:gce_recovered} shows the distribution of the recovered best-fit parameters across the same realizations. Each point represents the maximum-likelihood estimate from one individual realization. The contours in this figure are therefore constructed differently from the pooled likelihood contours of Fig.~\ref{fig:gce_bb}. Rather than combining the likelihood surfaces, we calculate the covariance matrix of the recovered best-fit points in the $(m_\chi,\log_{10}\langle\sigma v\rangle)$ plane. The contours are then defined by
\[
(\mathbf{x}-\bar{\mathbf{x}})^{T}\mathbf{C}^{-1}
(\mathbf{x}-\bar{\mathbf{x}})
=
\Delta\chi^{2},
\]
where $\mathbf{x}=(m_\chi,\log_{10}\langle\sigma v\rangle)^{T}$, $\bar{\mathbf{x}}$ is the mean of the recovered best-fit points, and $\mathbf{C}$ is their sample covariance matrix. We use $\Delta\chi^{2}=2.30$ and $6.18$ to define the two-parameter $1\sigma$ and $2\sigma$ covariance ellipses, respectively. These ellipses summarize the scatter and correlations of the recovered maximum-likelihood estimates, but do not capture non-Gaussian features of the underlying distribution. 

We further quantify the improvement in the recovered parameter constraints using the area of the $1\sigma$ covariance ellipse,
\[
A_{1\sigma}
=
\pi(2.30)\sqrt{\det\mathbf{C}},
\]
in the $(m_\chi,\log_{10}\langle\sigma v\rangle)$ plane. For \SMW, the ellipse area decreases from $11.90\,\mathrm{GeV\,dex}$ for the pre-LSST sample to $5.32\,\mathrm{GeV\,dex}$ for the post-LSST sample and $0.647\,\mathrm{GeV\,dex}$ for the ultimate sample. These correspond to reductions of $55\%$ and $95\%$, respectively, relative to the pre-LSST case. For \MWest, the area decreases from $9.32\,\mathrm{GeV\,dex}$ to $5.17\,\mathrm{GeV\,dex}$ and $0.764\,\mathrm{GeV\,dex}$, corresponding to reductions of $45\%$ and $92\%$, respectively. Defining the figure of merit as $\mathrm{FoM}=A_{1\sigma}^{-1}$, the ultimate dwarf sample improves the parameter-recovery figure of merit by factors of approximately $18.4$ for \SMW and $12.2$ for \MWest relative to the corresponding pre-LSST samples.

\begin{figure}
    \centering
    \includegraphics[width=\columnwidth]{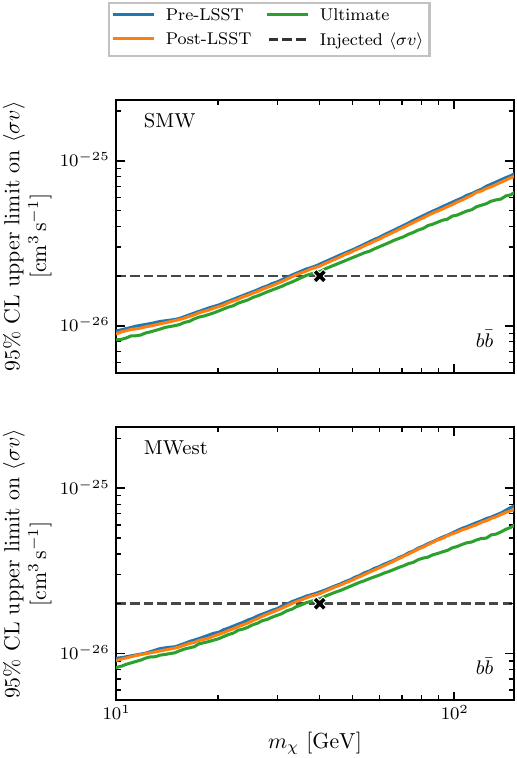}
    \caption{One-sided $95\%$ CL upper bounds on $\langle\sigma v\rangle$ as a function of dark-matter mass derived from simulated dwarf galaxy observations when the Galactic Center gamma-ray excess benchmark signal is injected (black marker). Limits from the pre-LSST (blue), post-LSST (orange), and ultimate (green) dwarf samples are shown using $17.8$ years of Fermi-LAT data. The top (bottom) panel shows results using the same \SMW (\MWest) catalogs from Fig.~\ref{fig:gce_recovered}.}
    \label{fig:gce_ul95}
\end{figure}

We also examine how the one-sided $95\%$ CL upper bound on $\langle\sigma v\rangle$ changes in the presence of the injected signal as the dwarf population increases, as shown in Fig.~\ref{fig:gce_ul95}. Using the same set of realizations as in Fig.~\ref{fig:gce_recovered}, we normalize the likelihood to its best-fit value at a given mass and determine the upper limit crossing at $2\Delta\ln\mathcal{L}=-2.71$ at that mass. The median across the full ensemble for each catalog and at each mass is plotted in Fig.~\ref{fig:gce_ul95}.  For both \SMW and \MWest, the pre-LSST and post-LSST upper bounds are nearly identical, while the ultimate dwarf sample yields a more noticeable, albeit modest, improvement across the mass range. As expected in the presence of a signal, the improvement in the upper limit is much more modest than the improvement in background only cases of Fig.~\ref{fig:ult_fermi} as well as for the reconstructed parameters shown in Figs.~\ref{fig:gce_bb} and \ref{fig:gce_recovered}.

\section{Summary and Conclusions}
\label{sec:conc}
In this work, we have investigated how future discoveries of Milky Way satellite galaxies could improve the sensitivity of gamma-ray searches for dark matter annihilation. We introduced \BabyBison, a Monte Carlo pipeline that combines simulated satellite populations, observational selection functions, and a stacked gamma-ray likelihood analysis. We validated the pipeline against the published six-year Fermi-LAT sensitivity using the same observed dwarf sample, then used two cosmological simulation suites to forecast the sensitivity of existing and future gamma-ray telescopes to dark matter annihilation, considering the pre-LSST, post-LSST, and complete Milky Way satellite galaxy populations.

Although LSST is expected to significantly increase the number of known Milky Way satellites, these discoveries do not necessarily lead to a significant improvement in dark matter sensitivity. The stacked limits are dominated by a small number of galaxies with large $J$-factors, so most newly discovered systems contribute relatively little statistical weight. Consequently, the anticipated LSST discoveries generally yield only modest improvements in the annihilation limits in typical realizations, generally below 20\%.  The distribution of possible gains nevertheless has a substantial tail: discovering a particularly favorable high-$J$ system can improve the sensitivity far more than the median forecast suggests. LSST's contribution to future dark matter annihilation searches thus depends strongly on the properties of the satellites it discovers, rather than simply on their number. 

Importantly, our projections for the pre- and post-LSST dwarf samples are restricted to galaxies with $M_V < 0$ and $r_{1/2} > 10$\,pc. Several candidate dwarf galaxies with smaller sizes and/or lower luminosities have nevertheless been discovered \citep{Smith:2024,Cerny:2026}. If these systems are dark matter dominated, they could have exceptionally large $J$-factors \citep{Crnogorcevic:2023ijs, Circiello:2025}. The resolution limits of current simulations and the limited observational data, however, make it difficult to reliably predict the properties of the Milky Way satellite population in this regime. These faint, compact systems are also particularly difficult to distinguish from dark-matter-free star clusters, since the predicted velocity dispersions in both cases are below the systematic floor of current spectroscopic observations \citep{Cerny:2026, Thibodeaux:2026}. The increased sensitivity of Rubin's LSST, the Euclid mission, and the Nancy Grace Roman Space Telescope makes it plausible that many such systems will be discovered in the coming decade. Improved statistical precision and control of systematic uncertainties with next-generation spectroscopic facilities, such as ViaSpec \citep{Via:2026} and 30-meter-class telescopes \citep{Johns:2012,Skidmore:2015,Gilmozzi:2007}, could also enable reliable estimates of their dark matter content. In this context, we also present forecasts for the complete modeled population of Milky Way satellite galaxies, removing our restrictions on $M_V$ and $r_{1/2}$. Upcoming data will also better constrain the underlying galaxy--halo connection \citep{Nadler:2024ims}, further reducing systematic uncertainties in $J$-factor predictions.

Under our assumptions, the full population of Milky Way satellites would allow Fermi-LAT to test dark matter with thermal-relic annihilation cross sections for masses as high as several hundred GeV. A future instrument such as the Advanced Particle-astrophysics Telescope (APT) could further extend this sensitivity, particularly at high masses. These forecasts illustrate the complementary roles of optical surveys, which identify the targets, and gamma-ray observations, which test their annihilation signals.

Finally, we used \BabyBison to evaluate the prospects for detecting and characterizing a benchmark dark matter signal motivated by the Galactic Center gamma-ray excess. For this benchmark, LSST discoveries modestly increase the probability of detection and improve estimates of the dark matter mass and annihilation cross section. A complete satellite census offers substantially greater gains in the precision of these measurements. Observations of dwarf galaxies could therefore provide an important independent test of dark matter interpretations of the Galactic Center excess and, if a signal were established, help determine the underlying particle properties. These prospects give us reason for optimism about the role of future dwarf satellite observations in revealing the nature of dark matter.

\section*{Acknowledgments}
The work of AG, IC and CK is supported in part by the U.S.~Department of Energy, Office of Science, Office of High Energy Physics under Award Number DE-SC0024693. The work of CK is also supported by the U.S.~National Science Foundation Growing Convergence Research award 2428507. DH is supported by the Office of the Vice Chancellor for Research at the University of Wisconsin–Madison with funding from the Wisconsin Alumni Research Foundation. ADW gratefully acknowledges support from the NASA Fermi Guest Investigator Program Cycle 9 No.\ 91201 and the National Science Foundation under Grants Nos.\ AST-2307126 and AST-2407526. This research is partially funded by a
generous gift from Charles Simonyi to the NSF Division of Astronomical Sciences. The
award is made in recognition of significant contributions to Rubin Observatory’s Legacy
Survey of Space and Time.

Fermilab is managed by FermiForward Discovery Group, LLC under Contract No. 89243024CSC000002 with the U.S.\ Department of Energy, Office of Science, Office of High Energy Physics. The United States Government retains and the publisher, by accepting the article for publication, acknowledges that the United States Government retains a non-exclusive, paid-up, irrevocable, world-wide license to publish or reproduce the published form of this manuscript, or allow others to do so, for United States Government purposes.

This work used data from the Symphony and Milky Way-est suites of simulations, hosted at \href{https://web.stanford.edu/group/gfc/gfcsims/}{\textbf{https://web.stanford.edu/group/gfc/gfcsims/}}, which were supported by the Kavli Institute for Particle Astrophysics and Cosmology at Stanford
University, SLAC National Accelerator Laboratory, and the
US Department of Energy under contract No. DE-AC02-76SF00515 to SLAC National Accelerator Laboratory.

\bibliography{apssamp}
\end{document}